\documentstyle[aps,prl,epsf,floats]{revtex}
\begin{document}
\twocolumn[\hsize\textwidth\columnwidth\hsize\csname@twocolumnfalse%
\endcsname
\title{Mean Field Theory of a Quantum Heisenberg
Spin Glass}
\author{Antoine Georges${}^{1}$, Olivier Parcollet${}^{1,2}$, 
and Subir Sachdev${}^{3}$}
\address{${}^{1}$CNRS - Laboratoire de Physique Th{\'e}orique, Ecole Normale
Sup{\'e}rieure, 24 Rue Lhomond 75005 Paris FRANCE\\
${}^{2}$Serin Physics Laboratory, Rutgers University, Piscataway, 
NJ 08854 USA\\
${}^{3}$Department of Physics, Yale University,
P.O. Box 208120,
New Haven, CT 06520-8120, USA}
\date{August 2, 1999}
\maketitle
\begin{abstract}
A full mean field solution of a quantum Heisenberg spin glass model is
presented in a large-$N$ limit. A spin glass transition
is found for all values of the spin $S$. 
The quantum critical regime associated with
the quantum transition at $S=0$, and the various regimes in the
spin glass phase at high spin are analyzed. The specific heat
is shown to vanish linearly with temperature. In the spin-glass 
phase, intriguing connections between the equilibrium properties of 
the quantum problem and the 
out-of-equilibrium dynamics of classical models are pointed out.
\newline
\end{abstract}]

The interplay between quantum effects and disorder in spin-glasses
have been a subject of great recent interest \cite{bhatt}. On the experimental side,
the strength of quantum fluctuations
can be continuously tuned by varying {\em e.g.} an applied transverse magnetic
field \cite{rosen}. Progress on the theoretical side, has followed
two different routes. From the higher dimensional end, mean field
solutions and effective Landau theories have been obtained\cite{millerhuse,rsy}
for quantum Ising and rotor spin glasses,
with a special focus on
the vicinity of the quantum-critical point where the
glass transition temperature is driven to zero.
In low dimensions \cite{dsf,young,mot}, it has been shown that the low $T$ physics is
controlled by rare events (``Griffiths-McCoy'' effects) at strong disorder
fixed points.

However, no established
mean-field theory of the experimentally important case of
quantum Heisenberg spin-glasses, with full $SU(2)$ symmetry,
is yet available. Unlike the rotor/Ising models above,
each site has non-trivial Berry phases which impose the spin
commutation relations, and this is expected to
place these models in a different universality class \cite{SachdevYe}.
Bray and Moore  \cite{BrayMoore}
pioneered the study of a
model of Heisenberg spins on a fully connected (`Sherrington-Kirkpatrick')
lattice of $\cal N$ sites.
In this letter, we report a full solution of this model, both in the
paramagnetic and the glassy phase,
when the spin symmetry group is extended from $SU(2)$ to $SU(N)$ and the
large-$N$ limit is taken. The Hamiltonian is
\begin{equation}\label{DefSachdevYe}
H = \frac{1}{\sqrt{{\cal N}N}} \sum_{i<j} J_{ij}
\vec{S}_{i}\cdot \vec{S}_{j},
\end{equation}
where the $J_{ij}$ are independent, quenched random variables
with distribution:
$P (J_{ij}) \propto e^{-J_{ij}^{2}/ (2J^{2})}$.
In an imaginary time path-integral formalism, the model is mapped onto
a {\it self-consistent single site} problem with the action \cite{BrayMoore,SachdevYe} :
\begin{equation}\label{ActionEffective}
S_{\text{eff}} = S_{B} + \frac{J^{2}}{2N} \int_{0}^{\beta} \!\!d\tau  d\tau'\,
Q^{ab} (\tau -\tau')
\vec{S}^{a} (\tau ) \cdot \vec{S}^{b} (\tau')
\end{equation}
with $\beta = 1/k_B T$, and
the retarded interaction $Q^{ab}(\tau-\tau')$ obeys the self-consistency condition :
\begin{equation}\label{CondCoherence}
Q^{ab} (\tau -\tau') =(1/N^2)
\left\langle \vec{S}^{a} (\tau ) \vec{S}^{b}(\tau') \right\rangle_{S_{\text{eff}}}.
\end{equation}
Here, $a,b=1,\cdots,n$ denote the replica indices (the
limit $n\rightarrow 0$ has to be taken later), and
$S_B$ is the Berry phase in the spin coherent state path integral.
For $N=2$ the problem remains of considerable
difficulty even in this mean field limit. In \cite{BrayMoore}, as well as in most subsequent
work, the {\it static approximation} was used (see however \cite{kopec}),
neglecting the
$\tau-$dependence of $Q^{ab}(\tau)$; this may be appropriate in some regimes but
prevents a study of the quantum equilibrium dynamics, in
particular,
in the quantum-critical regime. This imaginary time dynamics has
however been studied recently in a Monte Carlo simulation
with spin $S=1/2$ by Grempel and Rozenberg
\cite{Grempel1}, but their study was limited to the paramagnetic phase.
In our large-$N$ limit, the problem is exactly solvable and, as
explained below, this limit provides a good description of the
physics of the $N=2$ mean field model, as far as the latter is known.
We find that in the paramagnetic phase, at low $S$ (where the quantum fluctuations
are the strongest), the quantum critical regime is
a {\it gapless}  quantum paramagnet already studied in
\cite{SachdevYe,Slush} and radically
different from the paramagnet obtained in the classical regime
(at large $S$), in which a local moment behavior persists down to
the glass transition.
In the spin glass phase, various regimes are obtained as a function of 
temperature $T$. 
The thermodynamic properties and the dynamical response functions
are analyzed below. 
Most notably, the low $T$ specific heat is found to have a linear
$T$ dependence, a behavior commonly observed experimentally
in spin-glasses but not often realized in mean-field {\it classical} models.
Furthermore, the equilibrium dynamics of the quantum case reveals
intriguing connections with some known features of the out-of equilibrium
dynamics of classical glassy models, an observation already made in
\cite{GiamLedou} in a different context.

\begin{figure}[t]
\epsfxsize=3in
\centerline{\epsffile{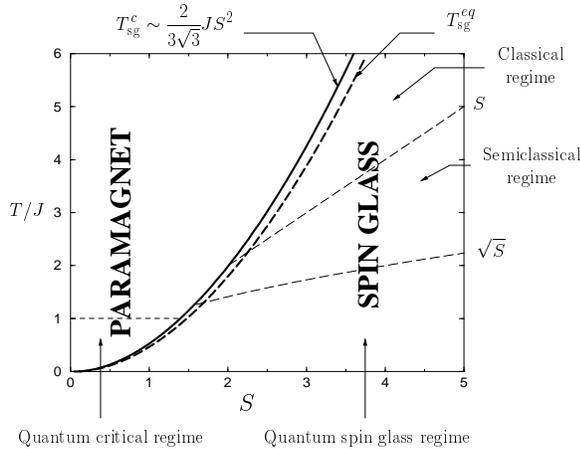}}
\caption{Mean-field phase diagram and crossovers of the large-$N$ quantum
Heisenberg spin glass (the various regimes are discussed in the text).}
\label{DiagrammePhase}
\end{figure}

To handle the large-$N$ limit, we use a Schwinger boson representation of the $SU(N)$
spin operators :  $S_{\alpha \beta}=
b^{\dagger}_{\alpha }b_{\beta } - S\delta_{\alpha \beta }$,
 corresponding to fully symmetric representations (one line of $NS$
boxes in the language of Young tableaux) where the number of bosons is constrained by
$\sum_{\alpha}b^\dagger_{\alpha}b_{\alpha}=NS$. In the $SU(2)$ case,
$S$ coincides with the usual definition of spin.
Fermionic representations can also
be considered but they actually do not lead to a spin-glass phase at any
temperature in the $N=\infty$ limit
\cite{SachdevYe}.
In the large-$N$ limit, the self-consistent single-site problem reduces to
a non-linear integral equation for the replicated boson
Green's function: $G^{ab} (\tau) \equiv  - \sum_{\alpha} \left\langle T b_{\alpha}^{a} (\tau)
b_{\alpha}^{\dagger b} (0) \right\rangle/N$ \cite{SachdevYe}:
\begin{eqnarray}\label{EqBase}
(G^{-1})^{ab} (i\nu_{n}) &=& i\nu_{n}\delta_{ab} +
\lambda^{a}\delta_{ab} - \Sigma^{ab} (i\nu_{n}) \\
\Sigma^{ab}(\tau ) &=& \ J^{2} \bigl( G^{ab} (\tau)\bigr)^{2}
G^{ab} (-\tau) \\
G^{aa} (\tau=0^{-} ) &=& - S
\label{EqBasex}
\end{eqnarray}
Here $\nu_n$ are the bosonic Matsubara
frequencies, and $G^{-1}$ stands for the inverse in replica space.
The (disorder-averaged) local spin correlation function is related to
$G^{ab}(\tau)$ by
$\chi_{\text{loc}}(\tau)\equiv \overline{\left\langle \vec{S}_i(0)
\cdot \vec{S}_i(\tau)\right\rangle} =
G^{aa}(\tau)G^{aa}(-\tau)$.
The resulting phase diagram, obtained by both analytical and
numerical studies of these equations, is displayed in
Fig. \ref{DiagrammePhase}, as a function of $S$
and $T/J$. Spin-glass ordering is found at any value of $S$.
The critical temperature increases as $JS^2$ at large $S$ (see below) and vanishes
in the limit $S\rightarrow 0$, which corresponds to the quantum critical
point in this model. Several crossovers are found {\it within} the
spin-glass phase, which will be described later.

We first describe the paramagnetic phase and the associated crossovers.
In this phase, the Green's function is replica diagonal
$G^{ab}(\tau)=G(\tau)\delta_{ab}$ and thus Eqs. (\ref{EqBase}-\ref{EqBasex}) reduce to
a single non-linear integral equation.
We emphasize that, as in any mean-field theory,
paramagnetic solutions of the mean-field equations can be found
even below the critical $T$ where an instability to ordering
occurs. At high $T$, we have nearly free
spins with an almost constant correlation function
$\chi_{\text{loc}}(\tau)\simeq S(S+1)$ and a Curie local susceptibility
$\chi_{\text{loc}}\equiv\int_0^{\beta}\chi_{loc}(\tau)d\tau
\simeq S(S+1)/T$. For large values of $S$, these
solutions smoothly evolve, as $T$ is reduced, into solutions which
still behave locally as local moments, but with a
Curie constant reduced by quantum fluctuations :
$\chi_{\text{loc}}\simeq S^2/T$.
This partial quenching occurs at a temperature of order $JS^2$ at
large $S$, of the same order but
{\it smaller} than the glass transition temperature.
These solutions actually have unphysical low-$T$ properties,
such as a divergent internal energy $U\simeq -J^2S^4/2T$ and a negative
entropy ($\propto -J^2S^4/4T^2$). These features are well-known in classical
mean-field models and simply signal the tendency to spin-glass
ordering. At smaller values of $S$ (Fig.\ref{DiagrammePhase}), a crossover to a
different kind of paramagnetic solution is found
below $T\simeq J$, where we enter the quantum critical
regime. In this gapless quantum paramagnet (`spin-liquid'),
investigated previously in \cite{SachdevYe,Slush},
the local response displays a scaling form for $\omega,T\ll J$,
$J\chi_{\text{loc}}''(\omega)\propto \tanh( \omega/2T)$, and the
local susceptibility diverges only logarithmically
$J\chi_{\text{loc}}\propto \ln (J/T)$.
In contrast to the local-moment solutions, this
paramagnet has finite residual low-temperature entropy \cite{SGlong},
so that the quenching of the entropy as $T$ is decreased takes place much
more gradually at low $S$ when quantum fluctuations are strong, than at
large $S$ in the classical regime. 
It can be shown analytically \cite{SGlong}
that these solutions of the mean-field equations exist down to
$T=0$ only for very low values of $S$, smaller than $S_c\simeq 0.05$.
For larger spins, a local-moment like solution is retrieved as $T$
is lowered below a temperature of order $JS^2$ (again below
the actual glass transition). 
However the spin-liquid solutions are
the relevant ones in the quantum critical regime {\it at finite
temperature}
$JS^2<T<J$ for an extended range of spin values which extend up to
$S\simeq 1$. The detailed analysis of the coexistence between these two
kinds of paramagnetic solutions at low $S$ is rather intricate and
will be presented elsewhere \cite{SGlong}.

In the Quantum Monte Carlo results of \cite{Grempel1} for the
paramagnetic phase of the $S=1/2$, $SU(2)$
model, the same reduction of the Curie constant from
$S(S+1)$ to $S^2$ was observed. Furthermore,
the relaxation function $\chi''(\omega)/\omega$ evolves
from a single peak of width $JS$ centered at $\omega=0$ to a three peak
structure in the
low-$T$ local moment regime. The central peak of weight $S^2$
corresponds to the residual local moment while two side peaks at an energy
scale $J^2S^3/T$ correspond to transverse relaxation \cite{Grempel1}. All these features
are captured by our solution in the large-$N$ limit, the only
qualitative difference being that no thermal broadening of the central
peak is found in this limit. Furthermore \cite{GrempelPrivate},
numerical results not reported  in \cite{Grempel1} reveal that in
a limited  intermediate $T$ range of the $SU (2)$ $S=1/2$ model, spin
liquid solutions similar to those found here in the 
quantum critical regime are observed. Although a logarithmic regime is not
directly visible in the $T$ dependence
of the local susceptibility because of this limited range,
quantum-criticality is directly apparent in a non-monotonic
$T$ dependence of the local spin correlation function 
$\chi_{loc}(\tau)$.

We now turn to the analysis of the spin-glass phase. We first note that
the spin-glass transition is {\it not} signalled
by the divergence of the spin-glass susceptibility (which is actually of
order $1/N$) \cite{susc}. In the ordered phase, the boson Green's function can be
parameterized as follows:
\begin{equation}
\label{AnsatzParisi}
G_{ab} (\tau ) = \left( \widetilde{G} (\tau ) - \tilde{g} \right)\delta_{ab}
 - g_{ab} (1-\delta_{ab})
\end{equation}
where $g_{ab}$ is a constant $n\times n$ matrix  and $g_{1}$ a constant,
fixed  so that  $\widetilde{G}$ is
regular at $T=0$, {\sl i.e. } $\widetilde{G} (\tau \rightarrow \infty)=0$.
The usual spin-glass order parameter
\cite{MezardParisiBook} is $q_{ab}=g_{ab}^2$.
We have searched for replica-symmetry
broken solutions by making a Parisi {\it Ansatz} for
$g_{ab}$ and found that {\it single-step} replica symmetry breaking
always applies \cite{kopec}.
 The Parisi function $g(x)$ associated with $g_{ab}$ is
thus piecewise constant:
$g(x)=0$ for $x<x_c$, $g(x)=g(1)=\sqrt{q_{EA}}\equiv g$ for
$x>x_c$, where $q_{EA}$ is the Edwards-Anderson order parameter;
this also implies that $\tilde{g} =g$.
For the following discussion, it is convenient to define the parameter
$\Theta\equiv -J\widetilde{G}(i\nu=0)/g$.
Using standard inversion formulas for a Parisi matrix \cite{MezardParisiManifold},
the full set of mean-field equations read:
\begin{eqnarray}\label{EqSG}
\left( \widetilde{G} (i\nu_{n})\right)^{-1} &=& i\nu_{n} - Jg/\Theta -
\left(\widetilde{\Sigma} (i\nu_{n}) - \widetilde{\Sigma} (0) \right)\label{EqSG_Dyson} \\
\nonumber
\widetilde{\Sigma} (\tau ) &\equiv&  J^{2} \left(
\widetilde{G}^{2} (\tau )\widetilde{G} (-\tau ) -
 2 g \widetilde{G} (\tau ) \widetilde{G} (-\tau)    \right.\\
&\quad& \left. -  g \widetilde{G}^{2} (\tau )  + 2 g^{2} \widetilde{G} (\tau )
+  g^{2} \widetilde{G} (-\tau)\right) \label{EqSG_DefSigTilde}\\
\widetilde{G} (\tau =0^{-}) &=&g  - S  \label{EqSG_NumberParticle}\\
\beta x_c &=&  \left(1/\Theta - \Theta  \right)/Jg^2
\label{EqSG_x}
\end{eqnarray}
However, these equations {\it do not determine }
$\Theta$ (or equivalently the breakpoint $x_c$)
as also happens in classical spin-glass model with a single-step of
replica symmetry breaking: there is a continuous family of solutions
parametrized by $\Theta$, which has to be determined by independent considerations.
Two possibilities have appeared in previous work:
({\em i}) Determine $\Theta $ by minimizing the
free energy, as a function of $\Theta$, or
({\em ii}) impose a vanishing lowest eigenvalue of the fluctuation
matrix  in the replica space (the ``replicon'' mode).
Criterion ({\em i}) is certainly the natural one from the
point of view of equilibrium thermodynamics. However, studies of
out-of-equilibrium dynamics of classical spin-glasses have revealed
\cite{CugliandoloKurchan} that
these lowest free-energy solutions can never be reached and that the
system ``freezes'' at a dynamical temperature $T_{\text{sg}}^{c}$,
given precisely by the onset of solutions satisfying the ``replicon'' criterion
({\em ii}).
In our quantum problem, both choices give
sensible solutions, but with entirely different spectra
of {\it equilibrium} dynamical fluctuations:
({\em i}) leads to a {\it gap} in $\chi''_{\text{loc}}(\omega)$, while
({\em ii}) is found to be the {\it unique} choice leading to a gapless spectrum.
A similar observation was made in the work of Giamarchi and Le Doussal
\cite{GiamLedou}
in their study of a one-dimensional quantum model with disorder. In the
present context, it seems natural to expect local gapless modes in the
ordered phase of a quantum spin-glass with continuous spin symmetry, and
these various considerations lead us to adopt ({\em
ii}).
Diagonalizing the fluctuation matrix in replica
space, we find the lowest eigenvalue
$e_1=3\beta J^2g^2(1-3\Theta^2)$ so that the replicon criterion leads
to  $\Theta=1/\sqrt{3}$, independent of $T$; the same value also
appears by independently imposing that $\widetilde{G}$ has a gapless
spectral weight.
In contrast, criterion ({\em i}) leads to
$2\ln\Theta+1/(4\Theta^2)+1/2-3\Theta^2/4=0$, or
$\Theta\simeq 0.44\dots$, and a gapped solution. We also note that the previous computation
shows that the replica symmetric solution $\Theta=1$ is unstable in
the spin glass phase. Moreover, it can be shown that it leads to
unphysical negative spectral weight at large-$S$.
Hence, a correct description of the
low-energy excitations of the quantum model requires replica
symmetry breaking at any finite $T$ in the spin-glass phase,
although the replica symmetry is restored at $T=0$ where
$x_c=0$ (from (\ref{EqSG_x})).

\begin{figure}[t]
\epsfxsize=2.8in
\centerline{\epsffile{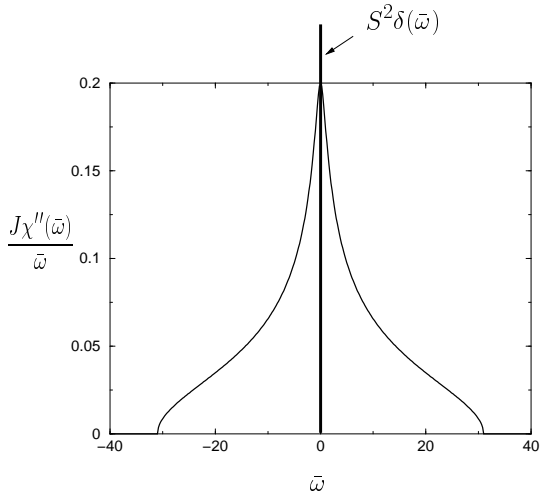}}
\caption{Relaxation function $\chi''(\omega)/\omega$ in the large-$S$ limit,  obtained from
(\protect\ref{EqQuartic}).}
\label{Kiom}
\end{figure}

Once $\Theta$ is determined, a full numerical solution of the above
equations can be performed. In particular, the ``equilibrium'' spin glass temperature
$T_{\text{sg}}^{eq}$ obtained from criterion ({\em i}) is {\it lower} than the
``dynamical'' transition temperature $T_{\text{sg}}^{c}$ obtained from
criterion ({\em ii}) (see Fig. \ref{DiagrammePhase}): this is not obvious a
priori, but is certainly required in our interpretation.
Further analytical insight can be
obtained in the limit of large $S$. This limit can actually be taken
in two distinct ways, revealing two crossovers within the spin-glass
phase displayed in Fig.\ref{DiagrammePhase}. If we take $S\rightarrow\infty$ while
keeping $T/JS^2$ fixed (i.e staying close to the critical temperature), all
non-zero Matsubara frequencies can be neglected (the static approximation
is accurate). In this limit, we find in particular
$T_{\text{sg}}^{c}\sim 2JS^{2}/3^{3/2}$.
Alternatively, keeping $\bar T=T/JS$ and $\bar \omega =\omega /JS$
fixed, we access the ``semi-classical'' regime of the spin-glass phase. In this limit,
the Green's function obeys a scaling
form $\widetilde{G}(\omega,T)= f(\bar \omega )/ (JS)$, where  $f$ turns out to be
independent of $\bar T$ and satisfies  :
\begin{equation}\label{EqQuartic}
f (\bar \omega )^{-1} = \bar \omega   - 1/\Theta - 3\Theta  -
f  (\bar \omega )- f^{\ast} (-\bar \omega )
\end{equation}
Eliminating $f^{\ast} (-\bar \omega )$ leads to a quartic equation
for $f (\bar \omega )$
on which all the above properties can be checked more explicitly.
A plot of the (gapless) relaxation function in the spin-glass phase
$\chi''(\bar \omega)/\bar \omega$ obtained from (\ref{EqQuartic}) is displayed in
Fig.\ref{Kiom}.

\begin{figure}[t]
\epsfxsize=2.8in
\centerline{\epsffile{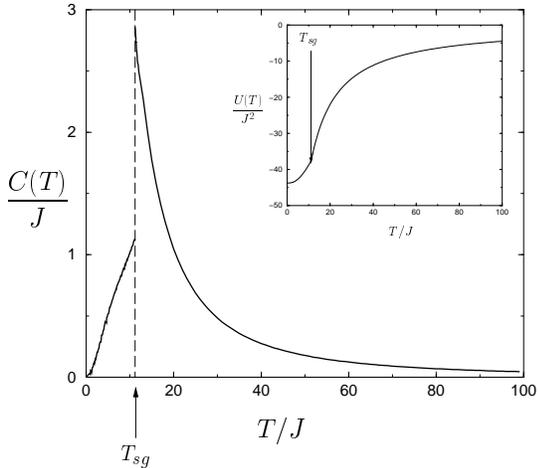}}
\caption{Specific heat $C (T)$ and internal energy $U (T)$ (inset) vs. 
temperature $T$, from a numerical solution of
Eqs. (\protect\ref{EqSG}-\protect\ref{EqSG_x}) for $S=5$.%
}
\label{CetU}
\end{figure}

Finally, we briefly describe the thermodynamic properties,
focusing on the $T$ dependence of the specific heat.
Numerical results for this quantity for intermediate spin are displayed on Fig \ref{CetU}.
They have been obtained from the $T$-derivative of the internal energy
$U= - J^{2}/2 \int_{0}^{\beta } G_{ab}(\tau )^{2}G_{ab}(-\tau )^{2} d\tau$,
where $G$ is a numerical solution of Eqs.(\ref{EqSG}-\ref{EqSG_x}).
Furthermore, a large-$S$, low$-T$ expansion of $U (T)$ can be done analytically and
leads to \cite{SGlong}:  
$U (T) = U (0) + a S \bar T ^{4} +b \bar T^{2} + \dots$
where  $a$ and $b$ are positive numerical coefficients.
Hence in the quantum regime defined by $T<J\sqrt{S}$ (see
Fig. \ref{DiagrammePhase}), the specific heat depends
{\it linearly on temperature}. Moreover, this behavior actually holds
numerically for intermediate values of the spin as displayed in Fig \ref{CetU}.

Despite being formulated over two decades ago\cite{BrayMoore}, a complete
understanding of the quantum Heisenberg spin glass at the mean-field 
level has proven elusive. Here we
have obtained a complete solution in a large $N$ limit,
and presented evidence that global aspects of the phase diagram
pertain also to the physical
$SU(N=2)$ case.
We described
crossovers in the vicinity of a quantum critical point accessed by varying the spin $S$,
but we can expect that some features and intermediate temperature
regimes will survive when it is accessed by varying other
parameters in the Hamiltonian, including doping with metallic carriers as in Kondo
lattice models \cite{msg,Slush}.
We have
also described the $T \rightarrow 0$ thermodynamics and spectral functions
within the spin-glass phase, which is something not previously
analyzed in any mean-field quantum spin glass model: we found a specific
heat linear in temperature, and a dynamical susceptibility
$\chi'' (\omega) /\omega \rightarrow
\mbox{const}$ as $\omega \rightarrow 0$.

We thank G.~Biroli, L.~Cugliandolo, T.~Giamarchi, D.~Grempel, 
P. Le Doussal, G.~Kotliar, M.~Rozenberg and A.~Sengupta 
for useful discussions. Laboratoire de Physique Th{\'e}orique de l'ENS is
UMR 8549 associ{\'e}e au CNRS et {\`a} l'Ecole Normale Sup{\'e}rieure.
S.S. was supported by NSF Grant No DMR 96--23181. 
A.G and O.P also acknowledge support of a NATO collaborative research grant.

{\it Note added}: It has been recently proven by one of us \cite{subir} 
that the behaviour $J\chi_{loc}''(\omega)\propto \mbox{const.}$ 
found above in the  
quantum critical regime also holds for the $SU(2)$ case.


\begin{references}
\bibitem{susc} This is also clear from the manner in which the
spin-glass order vanishes, with $x_c \rightarrow 1$ and $g$
finite \protect\cite{kopec}.
\end{references}

\end{document}